\documentclass[aps,twocolumn,showpacs,superscriptaddress,citeautoscript,prb,floatfix, 10pt]{revtex4-2}

\usepackage{bm}
\usepackage{graphicx}
\usepackage{amsmath,amssymb}
\usepackage{amsfonts}
\usepackage{color}
\usepackage{siunitx}
\usepackage{comment}

\definecolor{Red}{rgb}{0.9,0.3,0.3}

\begin{document}

\title{Dissociation of one-dimensional excitons by static electric field}

\author{Adriana Garc\'{\i}a}
\affiliation{GISC, Departamento de F\'{\i}sica de Materiales, Universidad Complutense, E-28040 Madrid, Spain}

\author{Alexander L\'{o}pez}
\affiliation{GISC, Departamento de F\'{\i}sica de Materiales, Universidad Complutense, E-28040 Madrid, Spain}

\author{Jorge Quereda}
\affiliation{2D Foundry Group. Instituto de Ciencia de Materiales de Madrid (ICMM-CSIC), Madrid, E-
28049, Spain}

\author{Francisco Dom\'{\i}nguez-Adame}
\affiliation{GISC, Departamento de F\'{\i}sica de Materiales, Universidad Complutense, E-28040 Madrid, Spain}

\begin{abstract}

The quantum states of an electron-hole pair in one-dimensional semiconductors under a static electric field are theoretically analyzed using a two-band model with on-site Coulomb interaction. In the absence of static field, the electron and hole are always bound, forming an exciton regardless of the Coulomb interaction strength, in contrast to what occurs in higher-dimensional semiconductors. The static field modifies the wave function of the electron-hole pair, turning bound states into continuum states. However, at low static fields, the linear optical spectra resemble those of the unbiased semiconductor, exhibiting a quadratic redshift of the main exciton absorption line as the field increases. When the static field exceeds a critical threshold, the exciton dissociates and the linear optical spectra exhibit signatures of the Wannier-Stark ladder with squally spaced peaks, making them a valuable tool for experimentally probing exciton dissociation.

\end{abstract}

\maketitle

\section{Introduction} \label{sec:Intro}

Inorganic semiconductor nanowires~\cite{Barth2010,Adame2019}, semiconducting carbon nanotubes~\cite{Yoshida2014,Perebeinos2005,Perebeinos2007}, molecular complexes and aggregates~\cite{Kobayashi1996,Amerongen2000} and conjugated polymers~\cite{Springborg2006} are quasi-one-dimensional gapped materials with a sizable concentration of free carriers (electrons and holes), opening opportunities for the design of novel optoelectronic devices. In addition to their potential technological interest, quasi-one-dimensional materials show enhanced electron-electron interactions compared to their three-dimensional counterparts and serve as a fundamental testing ground for the theoretical understanding of electron dynamics in low-dimensional systems~\cite{Baeriswyl2005}.  

The optical and optoelectronic properties of low- dimensional conducting materials are largely determined by the generation and recombination of excitons. Once generated, excitons undergo relaxation processes~\cite{Lee2023}, and their stability and lifetime are strongly dependent on a number of factors, such as the magnitude of the bandgap, temperature, crystalline quality of the material or strain, to name a few. Among them, high local electric fields can induce the dissociation of the exciton by pulling apart the electron and the hole. Exciton dissociation can happen, for instance, in the vicinity of metallic contacts, where an abrupt potential drop is expected. 

In this work, we theoretically address exciton dissociation by a static electric field. In particular, we focus on the interplay between the static electric field and the electron-hole Coulomb interaction. The temperature will be assumed low enough so that the effects of thermal baths and other dephasing mechanisms can be neglected. After introducing the model Hamiltonian, we calculate the linear optical spectra of the system by varying the strength of the static electric field. The absorption peaks below the exciton band split into equally space peaks, signaling the occurrence of the Wannier-Stark ladder in the energy spectrum. The main absorption peak is quenched when the magnitude of the static field exceeds a critical value, indicating the dissociation of excitons. The critical field depends on the magnitude of the Coulomb interaction between the electron and the hole, pointing out that the dissociation of the exciton happens when the static field enables the transition to continuum states. Linear optical spectra exhibit significant changes below and above the critical field. Our results are in excellent agreement with those reported in narrow carbon nanotubes, providing further insight into the complex dynamics of excitons under strong static electric fields.

\section{Theoretical model}

We start with a one-dimensional, two-band Hamiltonian model that incorporates the Coulomb interaction between electrons and holes, as introduced in Ref.~[\citenum{Ishida1993}], and add the potential energy arising from a static electric field of magnitude $\mathcal{E}$ parallel to the chain. The model Hamiltonian divides as $H=H_e+H_h+H_\mathrm{int}$, where
\begin{align}
    H_{e} & =  \sum_{m} (Fm+\epsilon_{cv})\, a^{\dag}_{m}a_{m}^{}-t_\mathrm{e}\sum_{m}\left(a^{\dag}_{m+1}a_{m}^{}+\mathrm{H.c.}\right)\ ,\nonumber \\
    H_{h} & = -\sum_{m} Fm\, b^{\dag}_{m}b_{m}^{}-t_\mathrm{h}\sum_{m}\left(b^{\dag}_{m+1}b_{m}^{}+\mathrm{H.c.}\right)\ ,\nonumber \\
    H_\mathrm{int} & = - \sum_{m,n}U_{m-n}^{}a^{\dag}_{m}a_{m}^{}b^{\dag}_{n}b_{n}^{} \nonumber \\
    & \phantom{=}\,\, - V_\mathrm{d}\sum_{m}\left(a^{\dag}_{m+1}b^{\dag}_{m+1}a_{m}^{}b_{m}^{}+\mathrm{H.c.}\right)\ ,
    \label{eq:01}
\end{align}
where $F=e\mathcal{E}a$, $e$ and $a$ are the elementary charge and the lattice constant, respectively. We will take $a$ as the length unit hereafter. The operators $a_{m}$ and $b_{m}$, respectively, stand for the annihilation operator of an electron and a hole at the $m$th site, while $t_\mathrm{e}$ and $t_\mathrm{h}$ denote the corresponding transfer energies. We consider both $t_\mathrm{e}$ and $t_\mathrm{h}$ positive, consistent with a semiconductor that exhibits a direct band gap in the center of the Brillouin zone. The parameter $\epsilon_{cv}$ denotes the energy difference between the two orbitals from which the conduction and valence bands originate~\cite{Ishida1995}, and the bandgap of the semiconductor is $\epsilon_{cv}-2(t_\mathrm{e}+t_\mathrm{h})>0$. The term $\epsilon_{cv}$ will be omitted until stated otherwise. 
The long-range interaction between an electron and a hole is taken into account by the regularized Coulomb potential
%
%
\begin{equation}
    U_{m-n} = \left\{
    \begin{array}{cl}
    U_0 \ , & m=n\ ,\\
    U_1/|m-n|\ , & m\neq n\ ,
    \end{array}
    \right.
    \label{eq:02}
\end{equation}
with $U_0>U_1\geq 0$. In our calculations, we have verified that the effect of the long-range interaction does not qualitatively affect our main conclusions. For this reason, in this work we will restrict ourselves to presenting the results for $U_1=0$ and take $U_{m-n}=U_0\delta_{mn}$ in what follows. 
Finally, the last term in $H_\mathrm{int}$ with $V_\mathrm{d}>0$ arises from the electromagnetic dipole-dipole coupling between neighbor atoms. This interaction term is needed for the mobility of excitons since it induces the motion of the center of mass~\cite{Heller1951,Ishida1993,Ishida1995}.

Under the assumption of low exciton density, we consider only one electron-hole pair. For this reason, we can neglect the spin degree of freedom in the expression of the system's Hamiltonian given in eq. ~\eqref{eq:01}. Let $|g\rangle$ be the ground state (filled valence band and empty conduction band), with its eigenenergy set to zero. A single-exciton state can be expressed as
\begin{equation}
    |\nu\rangle = \sum_{m,n}\psi_{m,n}(\nu) \, a^{\dag}_{m} b^{\dag}_{n}\, |g\rangle\ ,
    \label{eq:03}
\end{equation}
with eigenenergy $E_{\nu}$. The Schr\"{o}dinger equation $H|\nu\rangle=E_{\nu}|\nu\rangle$ yields the corresponding equation for the amplitudes $\psi_{m,n}(\nu)$
\begin{align}
    E_\nu\psi_{m,n} & = F(m-n)\psi_{m,n}-t_\mathrm{e}\left(\psi_{m+1,n}+\psi_{m-1,n}\right)\nonumber \\
    & -t_\mathrm{h}\left(\psi_{m,n+1}+\psi_{m,n-1}\right)-U_0\delta_{mn}\psi_{m,n}\nonumber \\
    & -V_\mathrm{d}\delta_{mn}\left(\psi_{m+1,m+1}+\psi_{m-1,m-1}\right)\ ,
    \label{eq:04}
\end{align}
where the parametric dependence of the amplitudes on $\nu$ has been omitted for simplicity. 

Since the momentum of the center of mass is a constant of motion, the two-particle amplitude can be separated as follows
\begin{equation}
    \psi_{m,n} = \frac{1}{\sqrt{N}}\,e^{i(m+n)K-i\theta(K)\ell}\phi_{\ell}\ ,
    \label{eq:05}
\end{equation}
with $\ell = m - n = 0,\pm 1, \ldots$ and 
\begin{subequations}
\begin{align}
    e^{-i\theta(K)} &=\frac{1}{t(K)}\left(t_\mathrm{e} e^{iK}+t_\mathrm{h} e^{-iK}\right)\ , \nonumber\\
    t(K)&= \sqrt{t_\mathrm{e}^2+t_\mathrm{h}^2+2t_\mathrm{e} t_\mathrm{h} \cos (2K)} \ ,
    \label{eq:06a}
\end{align}
yielding
\begin{align}
    E_\nu\phi_{\ell} & = \Big[
       F\ell -\Big(U_0 + 2V_\mathrm{d}\cos (2K)\Big)\delta_{\ell 0} 
    \Big]\phi_{\ell}\nonumber \\
    & - t(K) \big(\phi_{\ell + 1}+\phi_{\ell - 1}\big) \ .
    \label{eq:06b}
\end{align}
\label{eq:06}%
\end{subequations}
Notice that the amplitude $\phi_\ell$ depends on both $\nu$ and $K$. The equation for the amplitudes at $K=0$ will be relevant to study the linear optical response of the system. In this case, equation~\eqref{eq:06b} reduces to
\begin{equation}
    E_\nu\phi_{\ell}  = \left(
       F\ell - U\delta_{\ell 0} \right) \phi_{\ell} - T \big(\phi_{\ell + 1}+\phi_{\ell - 1}\big) \ ,
    \label{eq:07}
\end{equation}
where $T=t_\mathrm{e}+t_\mathrm{h}>0$ and $U=U_0+2V_\mathrm{d}$.

\section{Linear optical spectrum}

In the regime of first-order time-dependent perturbation theory in the coupling to the electromagnetic field with frequency $\omega$, the linear absorption spectrum $\rho(\hbar \omega)$ is given by summing up all possible transitions from the ground state $|g\rangle$ to all single-exciton states $|\nu\rangle$
\begin{equation}
    \rho(\hbar\omega) = \sum_{\nu} \big|\langle \,\nu \mid \widehat{\mu} \mid g\,\rangle \big|^2 \delta(\Delta\hbar \omega - E_\nu)  \ , 
    \label{eq:08}
\end{equation}
where constant prefactors will be omitted as they are not relevant to this analysis. Here we have defined $\Delta\hbar\omega=\hbar\omega - \epsilon_{cv}$.  The dipole operator $\widehat{\mu} = \sum_{m} \left(a_{m}^{\dag}b_{m}^{\dag}+a_{m}^{}b_{m}^{}\right)$ is given in units of the atomic transition dipole moment, which is assumed to be oriented parallel to the atomic chain. It is straightforward to calculate the oscillator strength $|\langle \,\nu \mid \widehat{\mu} \mid g\,\rangle |^2$ and to demonstrate that only states with $K=0$ are coupled to light. The absorption spectrum is then found to be
\begin{equation}
    \rho(\hbar\omega) \propto \sum_{\nu} \left|\phi_{0}(\nu)\right|^2 \delta(\hbar \Delta\omega - E_\nu)  \ , 
    \label{eq:09}
\end{equation}
where $\phi_{0}(\nu)$ is obtained after solving the equation~\eqref{eq:07} with the appropriate boundary conditions and imposing the normalization condition $\sum_\ell |\phi_{\ell}(\nu)|^2=1$. In practice, homogeneous or inhomogeneous broadening of optical absorption lines is taken into account by replacing $\delta(\hbar \Delta\omega - E_\nu)$ by a Lorentzian function with broadening parameter~$\Gamma$
\begin{equation}
    \delta_{\Gamma}(\hbar \Delta\omega - E_\nu) = \frac{1}{\pi}\,\frac{\Gamma}{(\hbar \Delta\omega - E_\nu)^2+\Gamma^2} \ .
    \label{eq:10}
\end{equation}

The absorption spectrum can be derived in a closed form in the two limiting cases discussed below.

\subsection{Non-interacting electron-hole pair}

In the absence of Coulomb and dipole-dipole interactions ($U=0$), the normalized eigenfunctions and eigenergies at $K=0$ are given as~\cite{Stey1973,Adame2010}
\begin{align}
    \phi_{\ell}(\nu)&=J_{\ell -\nu}(2F/T)\ ,\qquad \ell = 0,\pm 1,\ldots \nonumber \\
    E_\nu&=\nu F\ , \qquad \nu = 0,\pm 1, \ldots
    \label{eq:11}
\end{align}
where $J_n(x)$ denotes the Bessel functions of integer order. Therefore, the energy levels are equally spaced, $F$ being the level spacing. This energy spectrum is known as the Wannier–Stark ladder, characterized by the spatial localization of the eigenstates in a region of the order of $T/F$ in units of the lattice spacing. 

\begin{figure}[ht]
    \centering
    \includegraphics[width=0.8\columnwidth]{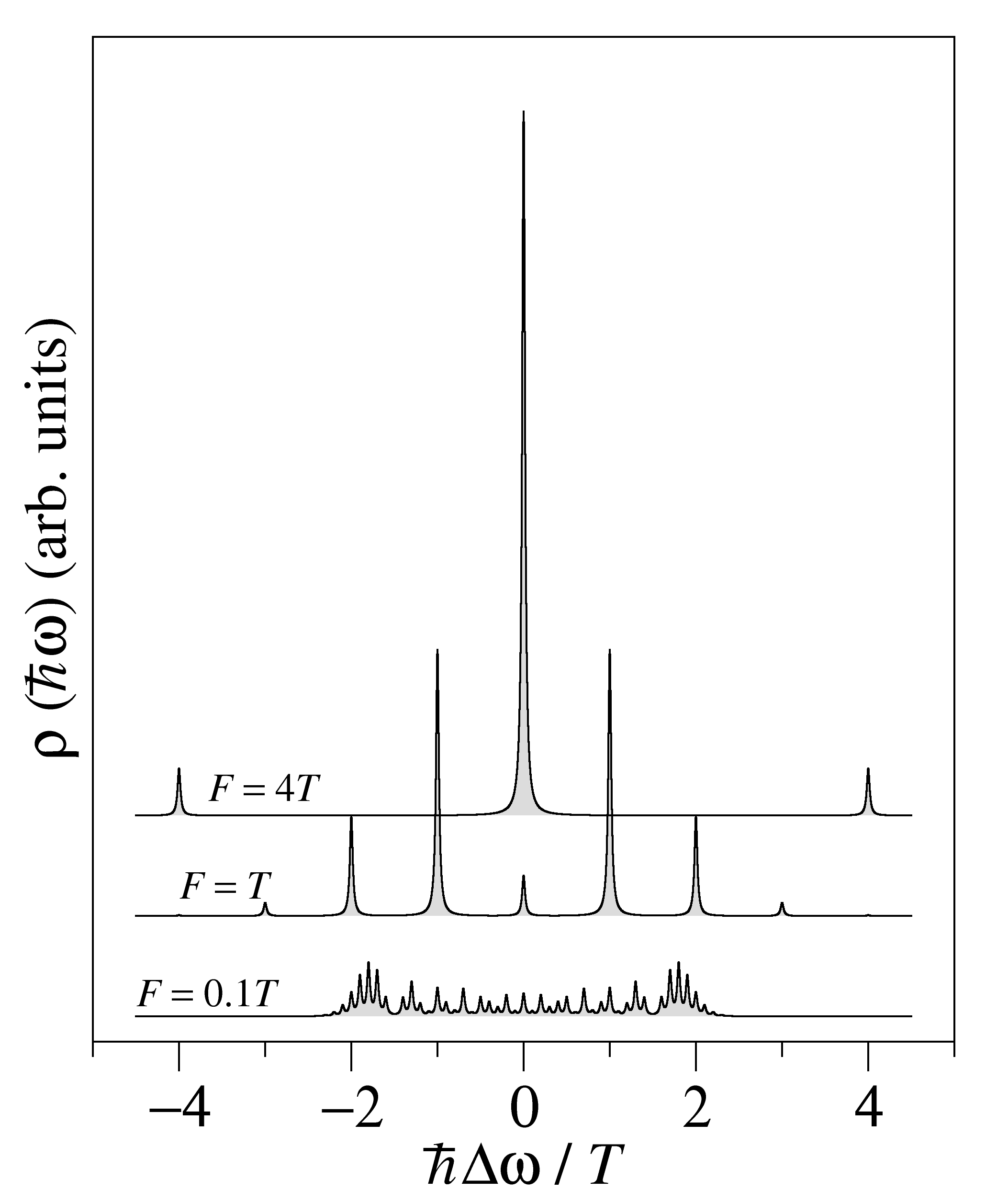}
    \caption{Absorption spectrum, in arbitrary units, in the absence of Coulomb and dipole-dipole interactions for different values of the static electric field, indicated on each curve. Spectra have been shifted upwards for clarity.}
    \label{fig:01}
\end{figure}

The optical spectrum~\eqref{eq:09} is given as
\begin{equation}
    \rho(\hbar\omega) \propto \sum_{\nu=-\infty}^{\infty} J_{\nu}^2(2T/F) \, \delta_{\Gamma}(\hbar \Delta\omega - E_\nu)  \ .
    \label{eq:12}
\end{equation}
Figure~\ref{fig:01} displays the absorption spectrum for different magnitudes of the electric field, with a broadening parameter $\Gamma = 0.02\,T$. At low field strengths, the states are only slightly perturbed and remain weakly coupled to light, resulting in a spectral width comparable to the bandwidth $4T$. As the electric field increases ($F=T$), clear signatures of the Wannier–Stark ladder emerge, characterized by equally spaced spectral lines whose intensity progressively decreases at higher values of $|\nu|$. In the limit of high electric field ($F>T$), the absorption spectrum is approximated up to the second order in $1/F$ as follows
\begin{align}
    \rho(\hbar\omega) &\propto \left(1-2\,\frac{T^2}{F^2}\right)\delta_\Gamma(\hbar \Delta\omega)\nonumber \\
    & +\frac{T^2}{F^2}\Big[\delta_\Gamma(\hbar \Delta\omega+F)+\delta_\Gamma(\hbar \Delta\omega-F)\Big] \ ,
    \label{eq:13}
\end{align}
as seen in Fig.~\ref{fig:01} for $F=4T$.

\subsection{Interacting electron-hole pair at zero field}

Another exactly solvable limiting case corresponds to an interacting electron-hole pair in the absence of static electric field. When $F=0$, there exist only a single localized solution of Eq.~\eqref{eq:07}. Labeling the localized state as $\nu=0$, we get
\begin{subequations}
\begin{align}
    \phi_{\ell}(0)&=\sqrt{\tanh \alpha}\, e^{-\alpha|\ell|}\ ,\nonumber \\
    e^{-\alpha}&=\frac{1}{2T}\left(\sqrt{U^2+4T^2}-U\right)\ . \nonumber\\
    E_0&=-\sqrt{U^2+4T^2}\ ,
    \label{eq:14a}
\end{align}
On the other hand, states inside the band are characterized by a wave number $q$ in the first Brillouin zone~\cite{Economou2006}
\begin{align}
    \phi_{\ell}(q) & =  \frac{1}{\sqrt{2\pi}}\,\frac{e^{iq\ell}}{1-\mathrm{i}\,U \left(4T^2-E_q^2\right)^{-1/2}}\ ,\nonumber \\
    E_q&=-2T\cos q \ .
    \label{eq:14b}
\end{align}
\label{eq:14}%
\end{subequations}

The optical spectrum is then given as 
\begin{align}
    \rho(\hbar\omega) &\propto  \tanh \alpha\,\delta_\Gamma(\hbar\Delta\omega - E_0)\nonumber \\
    +&\frac{1}{\pi} \frac{\sqrt{4T^2-\hbar^2\Delta\omega^2}}{U^2+4T^2-\hbar^2\Delta\omega^2}\,\Theta\left(2T-|\hbar\Delta\omega|\right)\ ,
    \label{eq:15}
\end{align}
where $\Theta$ is the Heaviside step function. To simplify the analysis, we consider the limit $\Gamma \to 0$ when evaluating the contribution of states within the band.

\begin{figure}[ht]
    \centering
    \includegraphics[width=0.75\columnwidth]{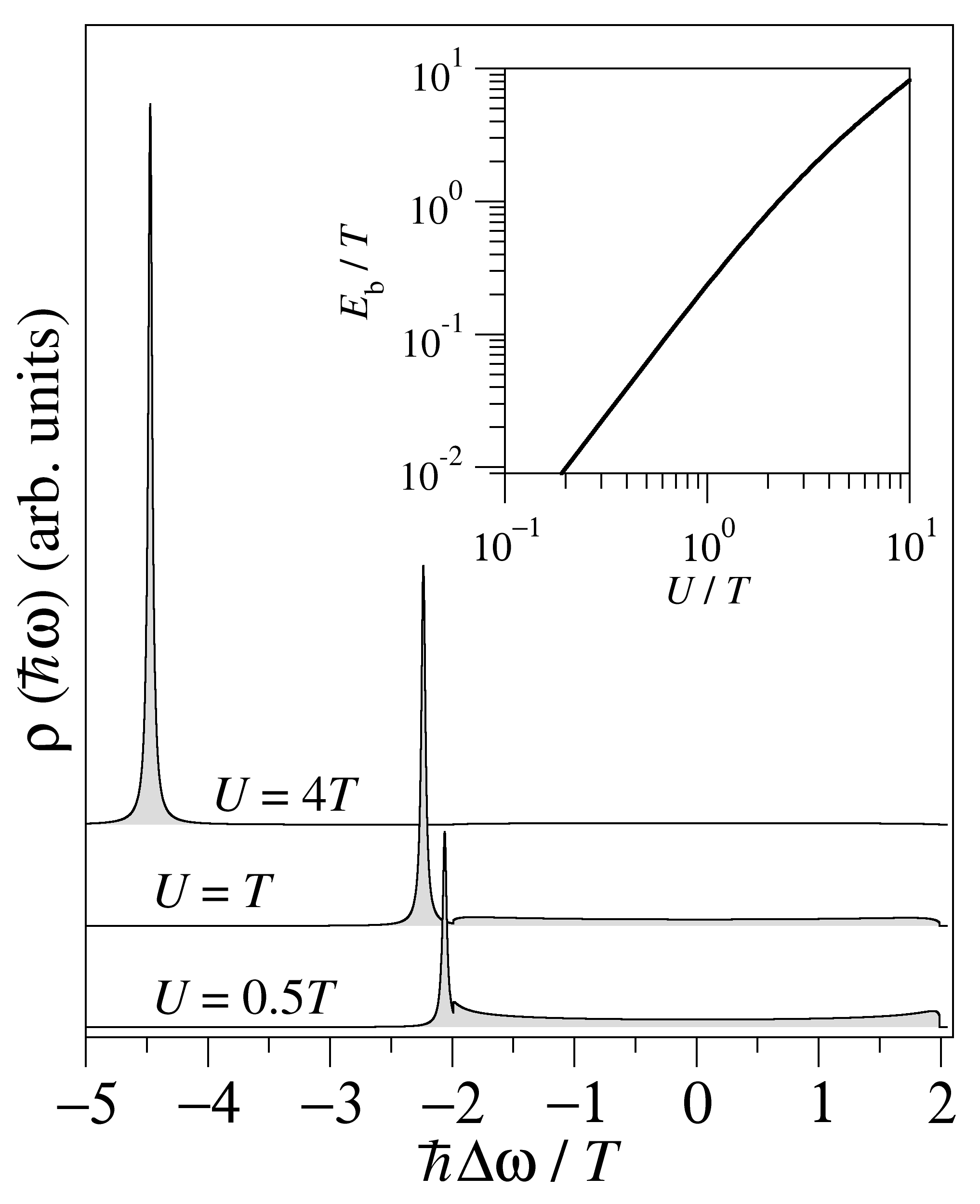}
    \caption{Absorption spectrum in arbitrary units in the absence of static electric field ($F=0$) for different values of $U$, indicated on each curve. In this plot, the optical bandgap corresponds to $\hbar\Delta\omega/T=-2$. The inset displays the binding energy $E_\mathrm{b}=-E_0-2T$ as a function of $U$.}
    \label{fig:02}
\end{figure}

Figure~\ref{fig:02} displays the absorption spectrum in the absence of a static electric field ($F = 0$), for various values of the interaction parameter $U$, with $\Gamma = 0.02\,T$. In the Wannier limit ($U < T$), the spectrum exhibits not only a prominent excitonic absorption line, but also a noticeable contribution from states within the band whose energies exceed the optical bandgap, corresponding to  $\hbar\Delta\omega=-2T$ in this plot. In contrast, this  contribution becomes negligible in the Frenkel limit ($U > T$), where only the excitonic line remains visible. The inset of Fig.~\ref{fig:02} shows that the exciton binding energy, defined as the energy difference between the lower band edge and the exciton level $E_\mathrm{b}=-2T-E_0$, increases with increasing $U$, according to Eq.~\eqref{eq:14a}.

\section{Interplay between Coulomb and external fields}

Equation~\eqref{eq:07} does not admit a closed-form analytical solution when both the static electric field and the Coulomb interaction between the electron and the hole are present simultaneously. Hence, numerical methods must be employed to compute the eigenfunctions and eigenvalues from which the absorption spectra can be derived. Rigid boundary conditions are imposed ($\phi_{\pm N}=0$) and the number of sites is set to $N=1000$ to ensure that finite-size effects are negligible.

\begin{figure}[ht]
    \centering
    \includegraphics[width=0.8\columnwidth]{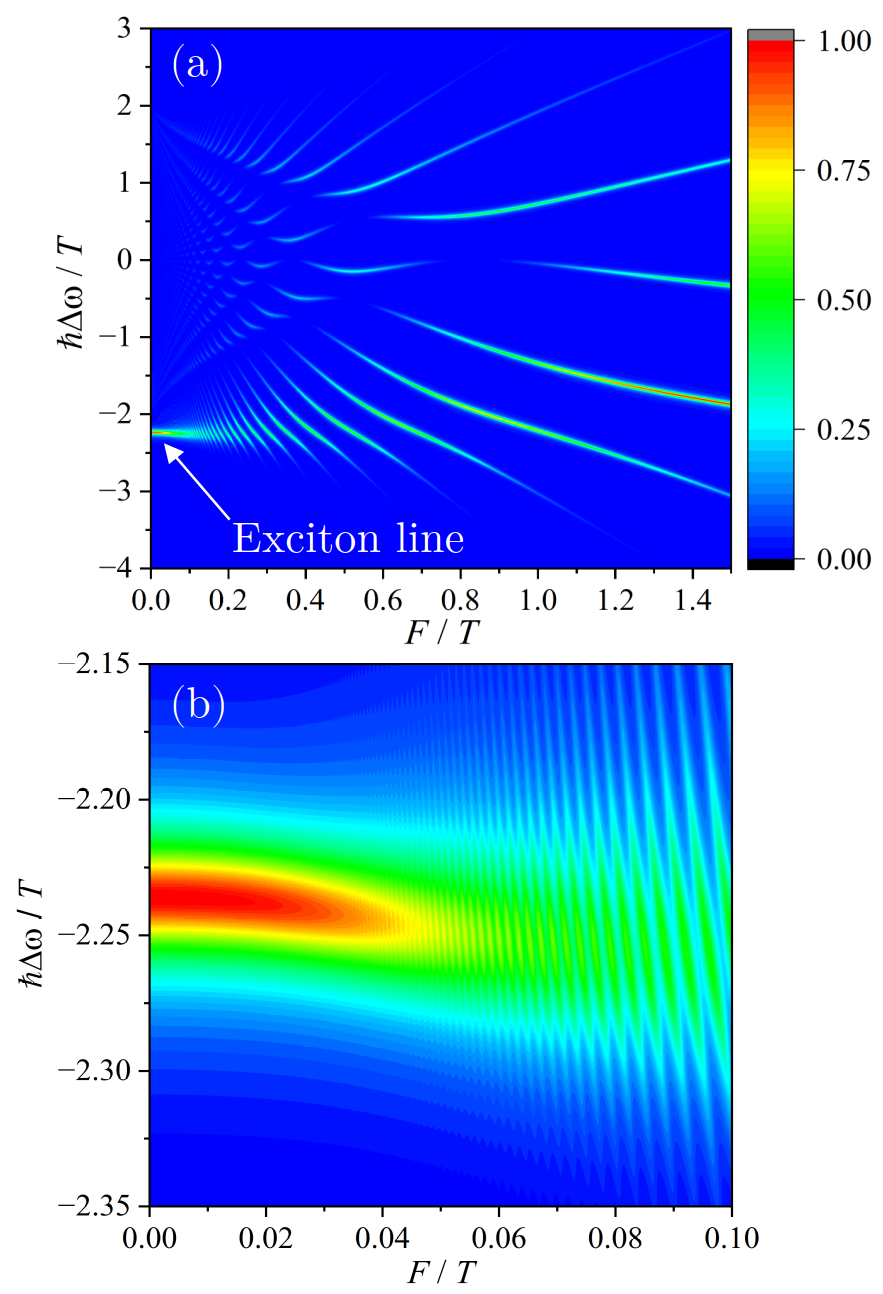}
    \caption{(a)~Absorption spectra as a function of energy $\hbar\Delta\omega$ and electric field $F$ for $U=T$, normalized to the maximum value of the dataset. The fan-chart of the Wannier-Stark ladder is revealed at high fields. (b)~Detailed view at low field strengths, illustrating the quadratic Stark effect and the progressive attenuation of the exciton line.}
    \label{fig:03}
\end{figure}

Figure~\ref{fig:03}(a) shows the absorption spectra as a function of energy $\hbar\Delta\omega$ and electric field $F$ for $U=T$, normalized to the maximum value of the dataset. At high electric fields, the fan-like absorption spectra map directly the Wannier-Stark ladder, in agreement with the results shown in Fig.~\ref{fig:01}. On the other side, at low electric fields, the exciton line is clearly revealed below the lower band edge, in correspondence to the spectra presented in Fig.~\ref{fig:02}. Figure~\ref{fig:03}(b) shows an enlarged view of the spectra at low electric field. On one hand, the exciton line experiences a redshift as the magnitude of the static electric field increases. On the other hand, there is a clear competition between the Coulomb attraction binding the electron and the hole, and the external electric field that tends to separate them. This competition leads to a progressive weakening of the excitonic line as the static electric field increases, signaling the dissociation of excitons.

\begin{figure}[ht]
    \centering
    \includegraphics[width=0.8\columnwidth]{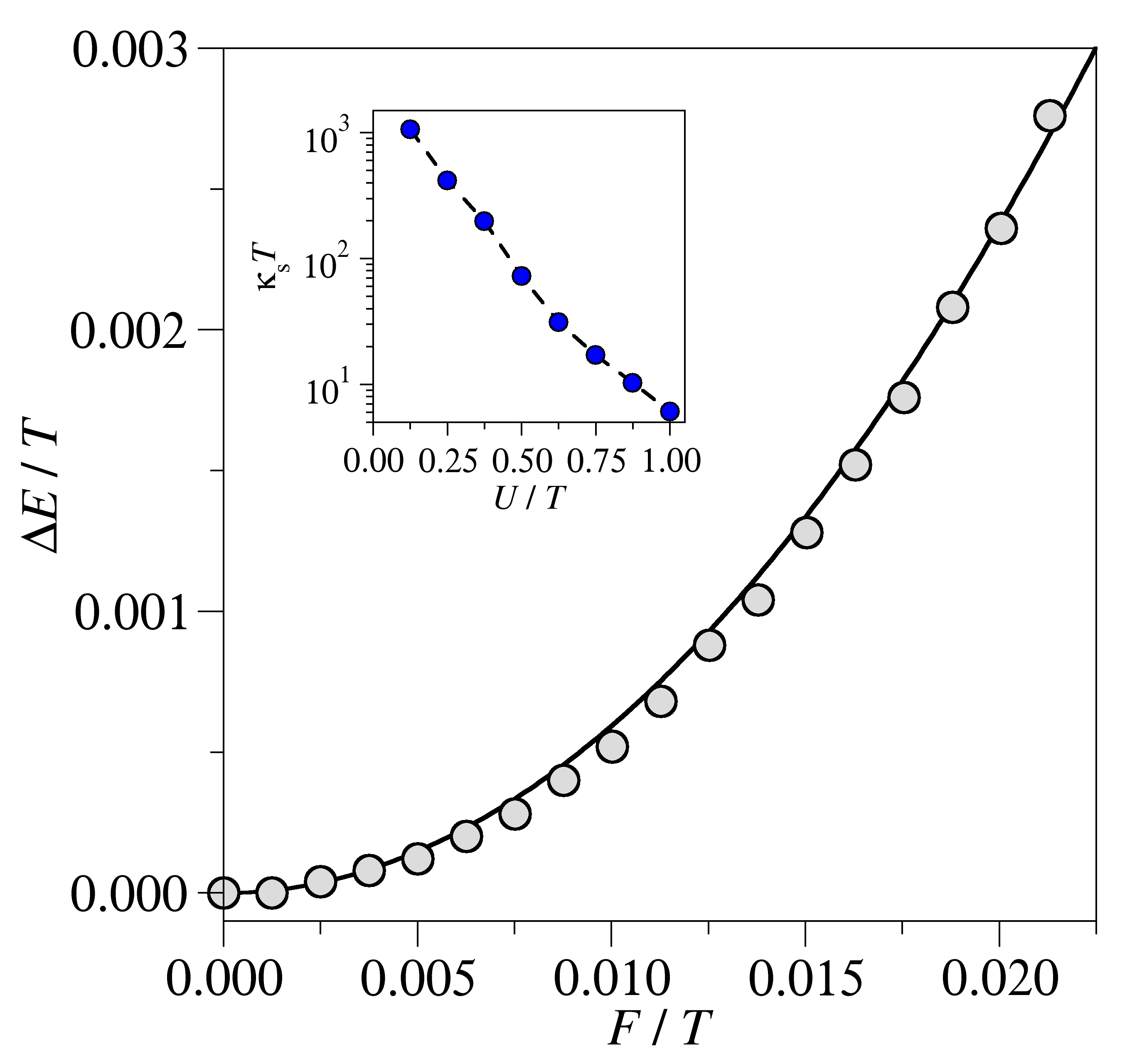}
    \caption{Stark shift of the exciton energy $\Delta E$ as a function of the electric field $F$ for $U=T$. Solid line correspond to the quadratic fitting $\Delta E=\kappa_\mathrm{s} F^2$. The inset shows the dependence of the coefficient $\kappa_\mathrm{s}$ on the interaction parameter $U$. Dashed line is a guide to the eye.}
    \label{fig:04}
\end{figure}

The excitonic line exhibits a redshift that increases quadratically with the strength of the static electric field due to the Stark effect~\cite{Perebeinos2007}, as seen in Figure~\ref{fig:03}(b). Figure~\ref{fig:04} shows the Stark shift $\Delta E$, defined as the absolute value of the change in exciton energy when an electric field is applied, for $U=T$. The quadratic dependence of the Stark shift on the static electric field can be cast as follows
\begin{equation}
    \Delta F = \kappa_\mathrm{s} F^2\ ,
    \label{eq:16}
\end{equation}
with $\kappa_\mathrm{s}>0$. This trend can be readily explained using second-order perturbation theory. The inset of Fig.~\ref{fig:04} shows that the parameter $\kappa_\mathrm{s}$ strongly depends on $U$ in a very pronounced way, exhibiting a sharp decrease as the on-site interaction increases. Therefore, the parameter $\kappa_\mathrm{s}$ decreases as the binding energy increases, which is in good agreement with the experimental results reported by Yoshida \emph{et al}. in air-suspended single-walled carbon nanotubes~\cite{Yoshida2014}.

\begin{figure}[ht]
    \centering
    \includegraphics[width=0.8\columnwidth]{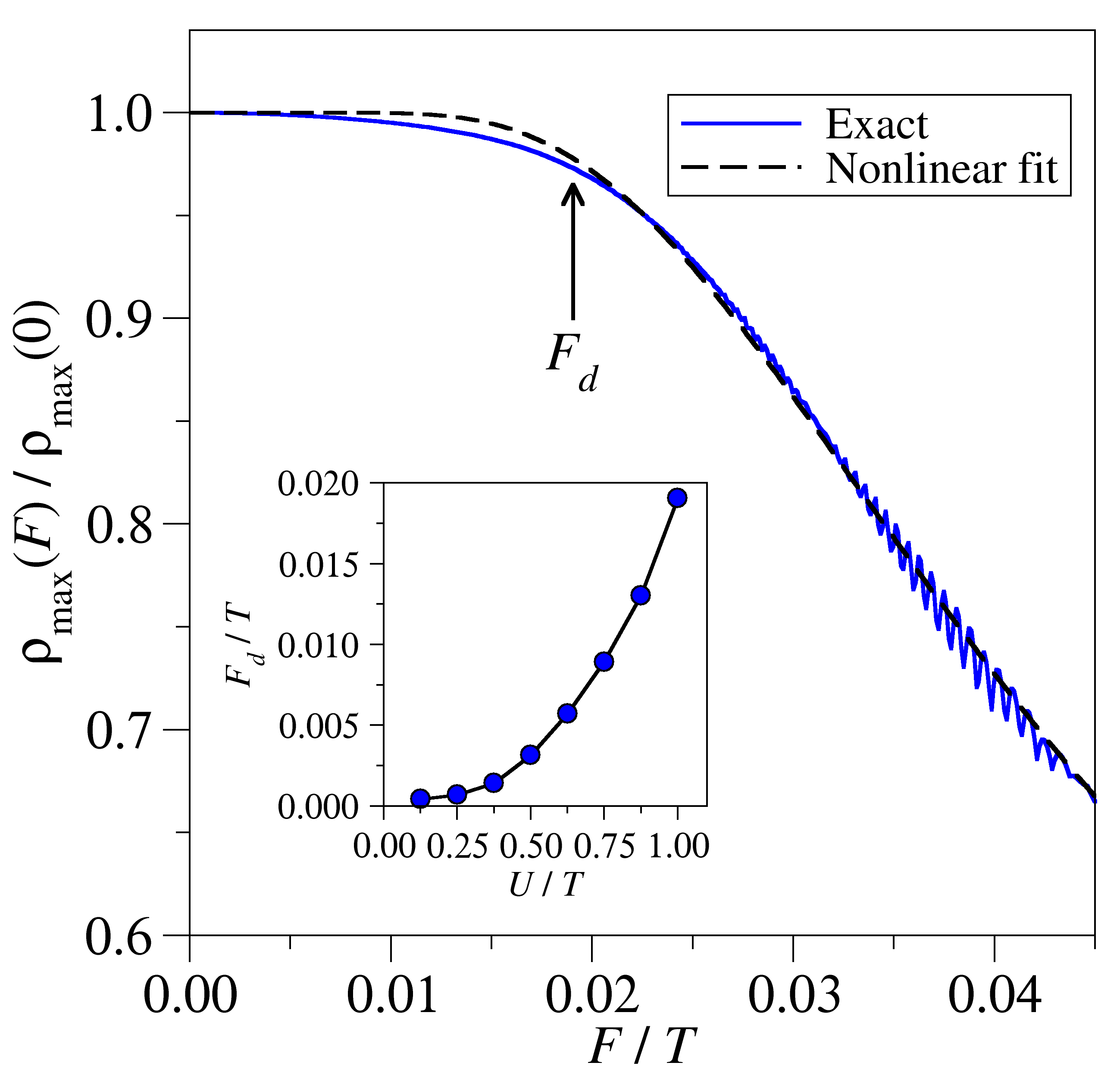}
    \caption{Maximum of the absorption coefficient of the exciton line $\rho_\mathrm{max}(F)$, normalized to its value at $F=0$, as a function of the static electric field for $U=T$ (solid blue line). Black dashed line corresponds to the nonlinear fit given by Eq.~\eqref{eq:17}. The inset shows the dependence of dissociation field $F_\mathrm{d}$ on the  interaction parameter $U$. Black solid line is a guide to the eye.}
    \label{fig:05}
\end{figure}

Figure~\ref{fig:05} shows the maximum of the absorption coefficient of the exciton line $\rho_\mathrm{max}(F)$, normalized to its value at $F=0$, as a function of the static electric field for $U=T$. At low static electric fields the exciton wave function is only slightly distorted, explaining the initial plateau up to $F\sim 0.015T$ seen in Fig.~\ref{fig:05}. According to the conventional description of Stark ionization, at intermediate electric field strengths, a potential energy barrier emerges due to the combined effects of the static electric potential and the Coulomb potential. In this regime, the exciton can dissociate via quantum tunneling. The exciton state becomes a resonance located below the barrier, corresponding to a relatively long-lived quasi-bound state. However, this state is significantly distorted, and its oscillator strength is notably reduced. This accounts for the reduction of the maximum of the absorption coefficient as the static electric field becomes stronger (see Fig.~\ref{fig:05}). Following the proposal by Najafov \emph{et al.} to describe the photoluminescence quenching of a phenyl-substituted poly phenylenevinylene derivative subject to an applied electric field~\cite{Najafov2006}, the electric-field dependence of the absorption coefficient depicted in Fig.~\ref{fig:05} can be well fitted with a one-dimensional model for tunneling through a triangular or parabolic potential barrier~\cite{Sze1981} 
\begin{equation}
    \rho_\mathrm{max}(F)=\frac{\rho_\mathrm{max}(0)}{1+ A \exp(-B/F)}\ ,
    \label{eq:17}
\end{equation}
where $A$ and $B$ are fitting parameters. The black dashed line in Fig.~\ref{fig:05} shows that the nonlinear fit is indeed accurate and reinforces the idea that exciton dissociation can be described as a tunneling process through a potential barrier. We estimate the magnitude of the static electric field required to induce dissociation, 
$F_\mathrm{d}$, as the value at which $\rho_\mathrm{max}(F)$ begins to decrease rapidly. Numerically, we find $F_d$ as the static field where the second derivative $\mathrm{d}^2\rho_\mathrm{max}(F)/\mathrm{d}F^2$ is maximal [see the arrow in Fig.~\ref{fig:05}]. The inset of Fig.~\ref{fig:05} shows that $F_\mathrm{d}$ increases as the on-site interaction potential $U$ increases, as expected.

\section{Comparison with first principle calculations}

Finally, a comparison is made between our predictions and the results obtained through more elaborate first- principle calculations in single-walled carbon nanotubes.
Perebeinos and Avouris have calculated the optical properties of semiconducting carbon nanotubes in an external static electric field directed along the tube axis using the Bethe-Salpeter equation solution ~\cite{Perebeinos2007}, starting from the single-particle spectrum of electrons and holes obtained within the GW approximation to the electron self-energy operator~\cite{Rohlfing2000}. 

To compare with our one-dimensional model, we focus our attention on the narrowest nanotube among those studied in ~\cite{Perebeinos2007}, that is, the $(13,0)$ tube for which the diameter is $d=1.03\,$nm. The exciton binding energy was found to be $E_\mathrm{b}=281\,$meV. The dissociation field was defined as the field in which the exciton linewidth becomes comparable to the exciton binding energy, yielding $\mathcal{E}_\mathrm{d}=84.5\,$V/$\mu$m.

The lattice parameter and the reduced effective mass in the $(13,0)$ tube are $a=0.246\,$nm and $m^{*}=0.033m_0$, respectively, where $m_0$ is the free-electron mass~\cite{Pedersen2004}. Therefore, the hopping energy of our model is found to be $T=\hbar^2/2m^{*}a^2=19.03\,$eV. We can estimate the on-site energy from the knowledge of the binding energy $E_\mathrm{b}=-2T-E_0=281\,$meV, where $E_0$ is given by Eq.~\eqref{eq:14a}, which yields $U=0.243T=4.63\,$eV. Finally, from the results presented in the inset of Fig.~\ref{fig:05} we get $F_\mathrm{d}=7\times 10^{-4}T=13.30\,$meV, leading to $\mathcal{E}_\mathrm{d}=\mathcal{F}_\mathrm{d}/(ea)=54\,$V/$\mu$m. The agreement with the value obtained from first-principles calculations is remarkable, especially considering that the definitions of the dissociation field are different.

\section{Conclusions}

The quantum states of an electron-hole pair in one-dimensional semiconductors under the influence of a static electric field are investigated theoretically within the framework of a two-band model that incorporates an on-site Coulomb interaction. In the absence of an external electric field, the electron and hole experience a strong Coulomb attraction due to the reduced screening in one-dimensional structures. As a result, they remain bound together to form a stable excitonic state. The application of a static electric field significantly alters the nature of the electron-hole pair wavefunction. As the field strength increases, it exerts a force that tends to pull the electron and hole apart, effectively modifying the potential landscape. This can lead to a transition from bound excitonic states to unbound, continuum-like states, particularly as the field becomes sufficiently strong to overcome the Coulomb attraction. Moreover, our model predicts a value for the critical field for this transition to occur that is in very good agreement with first-principles calculations.

Interestingly, at low field strengths, the linear optical absorption spectra remain qualitatively similar to those of the unbiased lattice, with the primary difference being a noticeable redshift of the exciton line. This redshift follows a quadratic dependence on the electric field strength and can be attributed to the Stark effect. As the electric field increases beyond a critical threshold, the exciton eventually dissociates. In this high-field regime, the linear optical spectra undergo a dramatic transformation, displaying a series of equally spaced absorption peaks known as the Wannier-Stark ladder. These peaks arise from the quantization of energy levels due to the presence of the uniform field and serve as a clear spectral signature of exciton dissociation. Consequently, the observation of Wannier-Stark features provides a powerful experimental means for probing the dynamics of field-induced exciton ionization in one-dimensional semiconductors.

\begin{acknowledgments}

Work at Madrid has been supported by Comunidad de Madrid (Recovery, Transformation and Resilience Plan) and NextGenerationEU from the European Union (Grant MAD2D-CM-UCM5) and Agencia Estatal de Investigación (Grant PID2022-136285NB-C31/3).

\end{acknowledgments}

\bibliographystyle{apsrev4-2}

\bibliography{references}

\end{document}